\documentclass[aps,prb,twocolumn,groupedaddress]{revtex4}
\usepackage{graphicx}
\begin{document}

\preprint{PRB:preprint}
\title{Low temperature ferromagnetism in (Ga, Mn)N}

\author{K. Sato}
\email{ksato@cmp.sanken.osaka-u.ac.jp}
\affiliation{The Institute of Scientific and Industrial Research, Osaka University, Osaka 567-0047, Japan}
\affiliation{Institut f\"ur Festk\"orperforschung, Forschungszentrum J\"ulich, D-52425 J\"ulich, Germany}

\author{W. Schweika}
\affiliation{Institut f\"ur Festk\"orperforschung, Forschungszentrum J\"ulich, D-52425 J\"ulich, Germany}

\author{H. Katayama-Yoshida}
\affiliation{The Institute of Scientific and Industrial Research, Osaka University, Osaka 567-0047, Japan}

\author{P. H. Dederichs}
\affiliation{Institut f\"ur Festk\"orperforschung, Forschungszentrum J\"ulich, D-52425 J\"ulich, Germany}

\date{\today}

\begin{abstract}
The magnetic properties of dilute magnetic semiconductors (DMSs) are calculated from first-principles by mapping the {\it ab initio} results on a classical Heisenberg model. By using the Korringa-Kohn-Rostoker coherent potential approximation (KKR-CPA) method within the local density approximation, the electronic structure of (Ga, Mn)N and (Ga, Mn)As is calculated . Effective exchange coupling constants $J_{\rm ij}$'s are determined by embedding two Mn impurities at sites i and j in the CPA medium and using the $J_{\rm{ij}}$ formula of Liechtenstein {\it et al.} It is found that the range of the exchange interaction in (Ga, Mn)N, being dominated by the double exchange mechanism, is very short ranged due to the exponential decay of the impurity wave function in the gap. On the other hand, in (Ga, Mn)As where $p$-$d$ exchange mechanism dominates, the interaction range is weaker but long ranged because the extended valence hole states mediate the ferromagnetic interaction. Curie temperatures ($T_{\rm{C}}$'s) of DMSs are calculated by using the mean field approximation(MFA), the random phase approximation(RPA)  and the, in principle exact, Monte Carlo method. It is found that the $T_{\rm{C}}$ values of (Ga, Mn)N are very low since, due to the short ranged interaction, percolation of the ferromagnetic coupling is difficult to achieve for small concentrations. The MFA strongly overestimates $T_{\rm{C}}$. Even in (Ga, Mn)As, where the exchange interaction is longer ranged, the percolation effect is still important and the MFA overestimates $T_{\rm{C}}$ by about 50-100\%.
\end{abstract}



\maketitle

Dilute magnetic semiconductors (DMSs), such as (In, Mn)As and (Ga, Mn)As discovered by Munekata {\it et al.} and Ohno {\it et al.}, have been well investigated as hopeful materials for spintronics \cite{matsukura}. Curie temperatures ($T_{\rm C}$'s) of these DMSs are well established \cite{matsukura,edmonds02,ku03} and some prototypes of spintronics devices have been produced based on these DMSs. The magnetism in these DMSs are theoretically investigated and it is known that the ferromagnetism in these systems as well as (Ga, Mn)Sb can be well described by Zener's $p$-$d$ exchange interaction, due to the fact that the majority $d$-states lies energetically in the lower part of the valence band \cite{ksato03}. Dietl {\it et al.} \cite{dietl02} and MacDonald {\it et al.} \cite{jungwirth02} successfully explained many physical properties of (Ga, Mn)As based on the $p$-$d$ exchange model, and first-principles calculations by Sato {\it et al.} showed that the concentration dependence of $T_{\rm{C}}$ in (Ga, Mn)As was well understood by the $p$-$d$ exchange interaction if a correction to the local density approximation (LDA) is simulated by the LDA+$U$ method with $U = 4$ eV \cite{ksato03}. 

While these $p$-$d$ exchange systems, in which the $d$-states of Mn impurities are practically localized, are well understood, there exist an even larger class of systems where the $d$-levels lie in the gap exhibiting impurity bands for sufficiently large concentrations. To these impurity band systems belong (Ga, Mn)N, (Ga, Cr)N, (Ga, Cr)As, (Zn, Cr)Te, (Zn, Cr)Se and many others as shown by first-principles calculations \cite{ksato02}. Most of these systems are controversially discussed in the literature, and an unambiguous determination of the ferromagnetism has only been reported for (Zn, Cr)Te with a relatively high Cr concentration of 20\% and a Curie temperature of 300 K \cite{saito03}. In particular in this class of materials, (Ga, Mn)N has been frequently mentioned as the most promising high-$T_{\rm{C}}$ DMS referring to the prediction of model calculations by Dietl {\it et al.} \cite{dietl02} and {\it ab-initio} results by Sato {\it et al.} \cite{ksato01}. Many groups have tried to fabricate ferromagnetic (Ga, Mn)N, but the experimental results are very controversial and confusing. After the first observation of the ferromagnetism of (Ga, Mn)N \cite{sonoda01}, many experiments followed, however reported $T_{\rm{C}}$'s are scattered between 20 to 940 K \cite{sonoda01,theodoropoulou01,overberg01,reed01,thaler02}. Moreover, recently Ploog {\it et al.} observed spin-glass behavior in 7\% Mn-doped GaN and suggested that the ferromagnetism observed in 14\% Mn-doped GaN originated from Mn-rich clusters \cite{ploog03}. Thus, the ferromagnetism in (Ga, Mn)N is still an open question which we reconsider in this letter. {\it Ab initio} calculations by Akai \cite{akai98} and others \cite{ksato03,ksato01,ksato02a,kulatov02,filippetti03,sanyal03} show that the magnetic properties of the above impurity band systems are dominated by double exchange mechanism and that the ferromagnetism is stabilized by the broadening of the impurity band. In the mean field approximation (MFA) high $T_{\rm{C}}$ values have been predicted ({\it e.g.}, 350 K for (Ga, Mn)N with 5\% of Mn, 500 K for (Ga, Cr)N with 5\% of Cr, 400 K for (Zn, Cr)Te with 5\% of Cr and so on) and the $\sqrt c$-dependence of $T_{\rm{C}}$ on concentration $c$ has been explained by band broadening \cite{ksato03,ksato02a}. Similar high, though slightly smaller, $T_{\rm{C}}$ values have also been obtained in the random phase approximation (RPA). 

In this paper, we will show that a general obstacle for ferromagnetism exists in these dilute systems, in particular in (Ga, Mn)N. Due to the large band gap the wave function of the impurity state in the gap is well localized, leading to a strong, but short ranged exchange interaction, being dominated by the nearest neighbors. Therefore, for low concentrations, the percolation of a ferromagnetic cluster through the whole crystal cannot be achieved, so that a ferromagnetic alignment of the impurity moments cannot occur. Thus a paramagnetic or disordered, spin-glass like, state is observed, in particular for low concentrations.

The electronic structure of DMS is calculated based on the local density approximation (LDA) by using the Korringa-Kohn-Rostoker (KKR) method. In this paper we focus on (Ga, Mn)N and (Ga, Mn)As as typical examples for the double exchange and the $p$-$d$ exchange systems, respectively. In these systems, Mn impurities distribute randomly at Ga sites in the host semiconductor being described as (Ga$_{1-c}$, Mn$_{c}$)X, where $c$ is the Mn concentration and X refers to N or As. To describe the substitutional disorder, we use the coherent potential approximation (CPA). In this framework, all Mn impurities are equivalent and consequently we suppose a ferromagnetic alloy. It has already been shown that the magnetic properties of metallic ferromagnetic alloys are well described within the CPA \cite{akai93}. While the CPA describes the electronic structure in the mean field approximation, we go beyond this approximation and explicitly calculate the exchange interaction $J_{\rm{ij}}$ between two impurities at sites i and j, which are embedded in the ferromagnetic CPA medium. For the evaluation of $J_{\rm ij}$ we use the frozen potential approximation \cite{oswald85} and apply a formula by Liechtenstein {\it et al.} \cite{liechtenstein87}. According to this formula, the total energy change due to infinitesimal rotations of the two magnetic moments at site i and j is calculated using the magnetic force theorem, and the total energy change is mapped on the (classical) Heisenberg model $H = -\Sigma _{\rm i \neq \rm j} J_{\rm{ij}} \vec e_{\rm i} \vec e_{\rm j}$, where $\vec e_{\rm i}$ is a unit vector parallel to the magnetic moment at site i, thus resulting in the effective exchange coupling constant $J_{\rm{ij}}$. This approach is already employed to estimate magnetic interactions in DMSs by Turek {\it et al.} \cite{turek03} and Bouzerar {\it et al.} \cite{bouzerar03}. For the present KKR-CPA calculations, we use the package MACHIKANEYAMA2000 coded by Akai \cite{machikaneyama2000}. We assume muffin-tin potentials and use the experimental lattice constants of the host semiconductors \cite{wyckoff}. It has already been shown that the lattice relaxations in (Ga, Mn)N and (Ga, Mn)As are very small \cite{mirbt02,kronik02,sanyal03}. Zinc blende structures are assumed both for GaN and GaAs. In reality, GaN has Wurtzite structure. However, results for both structures are practically identical, because splitting of impurity bands due to symmetry lowering is small \cite{sanyal03} and disorder induced band width always overcomes the splitting. The angular momenta are cut off at $l = 2$ in each muffin-tin sphere. All calculations are performed for the neutral charge state of Mn, so that doping effects are not included.

\begin{figure}
\includegraphics{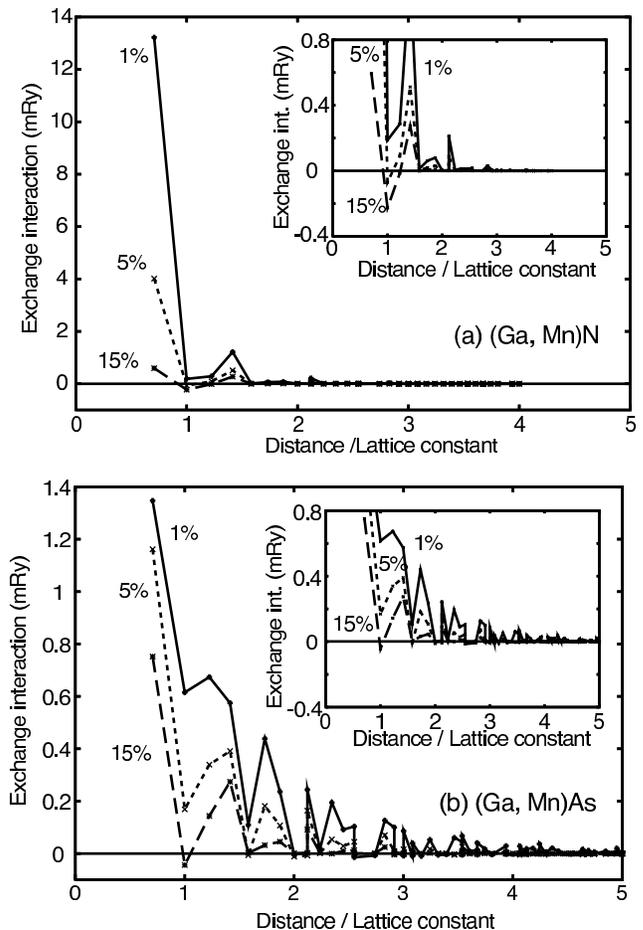}
\caption{\label{fig1} Calculated exchange interaction $J_{\rm{ij}}$ in (a) (Ga, Mn)N and (b) (Ga, Mn)As as a function of distance.}
\end{figure}

Fig.~\ref{fig1} shows the calculated exchange interactions $J_{\rm{ij}}$ in (Ga, Mn)N and (Ga, Mn)As. As shown in the fig.~\ref{fig1}-(a), in (Ga, Mn)N the interaction strength is strong, but the interaction range is short, so that the exchange coupling between nearest neighbors dominates. For example, nearest neighbor interaction $J_{\rm{01}}$ in 1\% Mn-doped GaN is about 13.5 mRy, while the other interactions are almost 2 orders of magnitude smaller than $J_{\rm{01}}$ except for $J_{\rm{04}}$. Therefore, in this case the very large mean field value of $T_{\rm{C}}$ is mostly determined by $J_{\rm{01}}$. For higher concentrations, $J_{\rm{01}}$ is suppressed and the interaction between next nearest neighbors becomes negative, resulting a complicated structure in the distance dependence of the exchange interaction. Concerning to the mechanism of the ferromagnetism, it has already been pointed out that the double exchange mechanism dominates in (Ga, Mn)N where pronounced impurity bands appear in the gap \cite{ksato02a,ksato02,ksato01,ksato03}. It is intuitively understood that the exchange interaction in (Ga, Mn)N becomes short ranged due to the exponential decay of the impurity wave function in the gap. In contrast to (Ga, Mn)N, the exchange interaction has long tails in (Ga, Mn)As in particular for low concentrations as shown in fig.~\ref{fig1}-(b). The qualitative difference in the interaction range between (Ga, Mn)N and (Ga, Mn)As is apparent from the figure. In (Ga, Mn)As the $p$-$d$ exchange interaction becomes important as shown in ref.~\cite{ksato03}. Since the extended hole state mediates the ferromagnetic interaction \cite{dietl02}, the interaction range is long ranged in $p$-$d$ exchange systems essentially. Actually the interaction extends farther than 3 lattice constants (20th shell). For higher concentrations, due to the screening of the pair interaction by the other impurities, interaction range becomes slightly shorter. 

As is well known the LDA predicts the position of localized d-levels at too high energy. However, according to recent calculations by Shick {\it et al.} \cite{shick04}, the LDA+U calculations only slightly affect the impurity bands at the Fermi level in (Ga, Mn)N due to the extended nature of the anti-bonding $t_{2}$ states of the impurity bands. Therefore, the LDA provides a fairly good description of the magnetic properties of (Ga, Mn)N. Even if the nearest neighbor interactions are changed in the LDA+U calculations, this will not affect much the Curie temperatures for low concentrations, because only the longer ranged interactions are relevant due to the percolation effects. On the other hand, as we have already shown in ref.~\cite{ksato03}, the LDA+U calculations with U=4eV yield a different description of the magnetism in (Ga, Mn)As. This effect could change the calculated $T_{\rm{C}}$ values slightly, however, the exchange interaction in (Ga, Mn)As still remains long ranged and the basic argument of the following discussion is not affected. 

\begin{figure}
\includegraphics{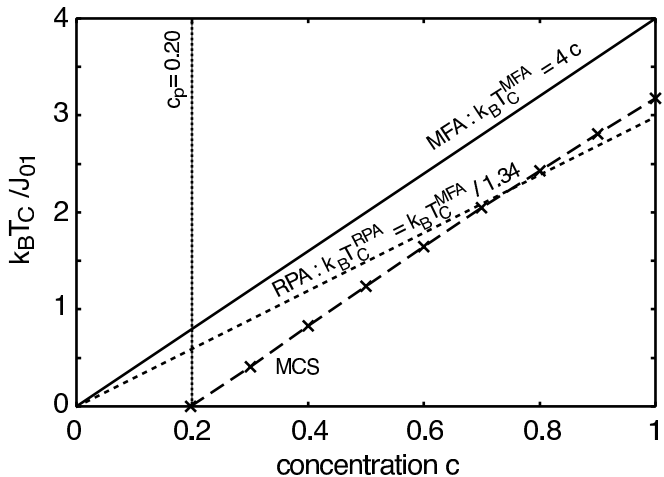}
\caption{\label{fig2} Curie temperatures of nearest neighbor Heisenberg model in fcc structure. $T_{\rm{C}}$'s are calculated by the mean field approximation (solid line), the random phase approximation (dotted line) and Monte Carlo simulation (crosses). The percolation threshold is 0.20 for the fcc structure.}
\end{figure}

It is well known that the Curie temperature in the mean field approximation $T_{\rm{C}}^{\rm{MFA}}$ is calculated as $k_{\rm{B}}T_{\rm{C}}^{\rm{MFA}} = \frac{2}{3} c\Sigma_{i \neq 0} J_{0i}$, where $k_{\rm{B}}$ is Boltzmann constant. As shown in this equation, evaluation of $T_{\rm{C}}^{\rm{MFA}}$ does not require any information on the interaction range, because only the sum of the coupling constants appears in the equation. This simplification leads to significant errors in the calculated $T_{\rm{C}}$ of a dilute system with low concentrations. This fact is easily understood by simple consideration and is known as the percolation problem \cite{stauffer}. Let us suppose a Heisenberg model with a ferromagnetic exchange interaction only between nearest neighbors (nearest neighbor Heisenberg model), and consider what happens when the system is diluted with non-magnetic sites. When the concentration of magnetic sites is 100\%, we have a perfect ferromagnetic network. Due to the dilution, the network is weakened, and for a concentration below a percolation threshold the ferromagnetism cannot spread all over the system leading to paramagnetic state since due to missing longer ranged interactions the moments can no longer align. Obviously this effect is not counted in the mean field equation for $T_{\rm{C}}$, because the dilution effect is included only as a concentration factor $c$ in the equation. In case of the nearest neighbor Heisenberg model, the percolation thereshold $c_{\rm{p}}$ for fcc structure is 20\% \cite{stauffer}. In real cases the exchange interaction could reach beyond the nearest neighbors and the percolation threshold might be lower. However, in this paper we are interested in the concentration range well below the nearest neighbor threshold $c_{\rm{p}}$. Therefore the exact $T_{\rm{C}}$ values could be much lower than the mean field values, in particular for the double exchange systems like (Ga, Mn)N where the exchange interaction is very short ranged (fig.~\ref{fig1}-(a)). 

 In order to take the percolation effect into account, we perform Monte Carlo simulations (MCS) for the effective classical Heisenberg model. The thermal average of magnetization $M$ and its powers are calculated by means of the Metropolis algorithm \cite{binder}. Due to the finite size of super cells used in the simulation, it is difficult to determine $T_{\rm{C}}$ from the temperature dependence of $\langle M(T) \rangle$. In particular, when considering dilute systems, finite size effects and appropriate finite size scaling are of particular importance for a correct and efficient evaluation of $T_{\rm{C}}$ by Monte Carlo simulations. To avoid this difficulty, we use the cumulant crossing method proposed by Binder \cite{binder}. This method uses the finite size scaling in the fourth order cumulant $U_{4}$ which is defined as $U_{4} = 1 - \langle M^{4} \rangle / (3 \langle M^{2} \rangle ^{2})$. $U_{4}$ is calculated for various cell sizes and plotted as a function of temperature. If the cell size is larger than the correlation length, it can be shown that the $U_{4}(T)$ curves for different sizes cross each other at three characteristic temperatures. Two of them are $T = 0$ and $T = \infty $, and the other is $T = T_{\rm{C}}$. We use 3 cell sizes ($6 \times 6 \times 6, 10 \times 10 \times 10$ and $14 \times 14 \times 14 $ conventional fcc cells) to carry out the cumulant crossing method for $T_{\rm{C}}$ calculations. For each temperature, we perform 240000 Monte Carlo steps per site using every 20-th step for averaging. 

First, as a pedagogical example we show the calculated $T_{\rm{C}}$ for the dilute fcc nearest neighbor Heisenberg model as calculated by MFA, RPA and MCS in fig.~\ref{fig2}. For MCSs for dilute systems, we take 20 different random configurations of magnetic sites for the ensemble  average. As shown in fig.~\ref{fig2}, it is found that both MFA and RPA give reasonable estimations of $T_{\rm{C}}$ for $c = 1$, with the RPA being closer to exact MCS result. It has been analytically shown that for this model MFA gives upper limit of $T_{\rm{C}}$ and RPA gives lower limit \cite{froehlich76}. However, for $c \le 0.7$, MCS results are below RPA values and in particular below the percolation thereshold ($c_{\rm p} = 0.20$) the Curie temperature vanishes: $T_{\rm C} = 0$. Thus the serious deficiency of both MFA and RPA in the dilute concentration range is evident. 

\begin{figure}
\includegraphics{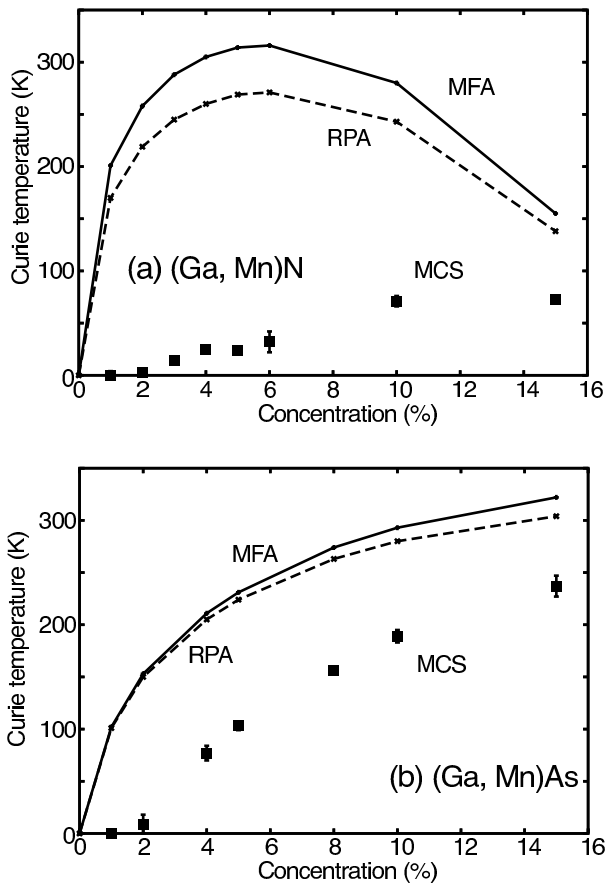}
\caption{\label{fig3} Curie temperatures of (a) (Ga, Mn)N and (b) (Ga, Mn)As calculated by the MFA (solid lines), the RPA (dotted lines) and the MCS (filled squares). For the MCS, the exchange interactions up to 15th shell are taken into account.}
\end{figure}

Next, we show the calculated $T_{\rm{C}}$ values of (Ga, Mn)N (fig.~\ref{fig3}-(a)) and (Ga, Mn)As (fig.~\ref{fig3}-(b)) as obtained by the MCS from the $J_{\rm{ij}}$ values in fig.~\ref{fig1}. Thirty configurations of Mn atoms are considered for averaging and $J_{\rm ij}$-interactions up to 15 shells are included; on the other hand, for the MFA and the RPA estimations interactions are included up to 72 shells. As shown in fig.~\ref{fig3}-(a), very small $T_{\rm{C}}$ values are predicted for low concentrations in (Ga, Mn)N. MFA and RPA values are almost 2 orders of magnitude too large. Thus we find that the magnetism is strongly suppressed due to the missing percolation of the strong nearest neighbor interactions. Only the weak, longer ranged interactions satisfy the percolation requirement, leading to small but finite Curie temperatures for 5, 10, and 15 \% of Mn. As shown in fig.~\ref{fig3}-(b), due to the longer ranged interaction in (Ga, Mn)As, the reductions from the MFA are not very large, but still significant. Naturally these changes are larger for smaller concentrations. The $T_{\rm{C}}$ values of 103 K obtained for 5\% Mn is in good agreement with the experimental values of 118 K reported by Edmonds {\it et al.} \cite{edmonds02}. This value refers to measurements in thin films which are free of Mn-interstitials representing double donors. Including interactions beyond the 15th shell, MCS could give slightly higher $T_{\rm C}$ values for low concentrations where the interactions do not converge within the 15th neighbors. At very high concentrations we expect our results to merge with the MFA and RPA values. 

In this Letter, we have shown by {\it ab-initio} calculations that (Ga, Mn)N shows no high-temperature ferromagnetism for low Mn concentrations. The strong ferromagnetic interaction of Mn nearest neighbor pairs does not become effective below the nearest neighbor percolation limit. The weak longer ranged interaction leads to a ferromagnetic phase with very low $T_{\rm{C}}$ of several tens Kelvin. Therefore the experimentally observed very high $T_{\rm{C}}$ values do not refer to a homogeneous ferromagnetic phase, but have to be attributed to small ferromagnetic MnN clusters and segregated MnN phases. Our results are of relevance for all DMS systems with impurity bands in the gap. To obtain higher Curie temperatures one needs longer ranged interactions and/or higher concentrations. The latter requirement naturally points to II-VI semiconductors, having a large solubility for transition metal atoms. The observation of a $T_{\rm{C}}$ value of 300 K for (Zn, Cr)Te with 20\% Cr \cite{saito03} is in line with these arguments. Similar results as presented above have been recently reported by a Swedish-Czech collaboration \cite{bergqvist04}

\begin{acknowledgments}
This research was partially supported by JST-ACT, NEDO-nanotech, a Grant-in-Aid for Scientific Research on Priority Areas A and B, SANKEN-COE and 21st Century COE from the Ministry of Education, Culture, Sports, Science and Technology. This work was also partially supported by the RT Network Computational Magnetoelectronics (Contract RTN1-1999-00145) of the European Commission.
\end{acknowledgments}


\end{document}